\documentclass[twocolumn,floatfix,prl]{revtex4}%
\usepackage[dvipdfmx]{graphicx}%
\usepackage{amsmath}%
\usepackage{amsfonts}
\usepackage{amssymb}
\usepackage{hhline}
\usepackage{bm}
\usepackage{mathrsfs}
\usepackage{color}

\def\e{{\epsilon}}
\def\k{{ {\bm k} }}
\def\p{{ {\bm p} }}
\def\q{{ {\bm q} }}

\def\0{{ {\bm 0} }}

\allowdisplaybreaks[4]

\begin{document}
\title{Emergence of Charge Loop Current in Geometrically Frustrated Hubbard Model:\\
Functional Renormalization Group Study}
\author{
Rina Tazai, Youichi Yamakawa and 
Hiroshi Kontani}

\date{\today }

\begin{abstract}
Spontaneous current orders due to odd-parity order parameters
attract increasing attention in various strongly correlated metals.
Here, we discover a novel spin-fluctuation-driven charge loop current (cLC)
mechanism based on the functional renormalization group 
(fRG) theory.
The present mechanism leads to the ferro-cLC order 
in a simple frustrated chain Hubbard model.
The cLC appears 
between the antiferromagnetic and $d$-wave superconducting ($d$SC) phases.
While the microscopic origin of the cLC
has a close similarity to that of the $d$SC,
the cLC transition temperature $T_{\rm cLC}$ can be 
higher than the $d$SC one for wide parameter range.
Furthermore, we reveal that
the ferro cLC order is driven by the strong enhancement of the 
forward scatterings $g_2$ and $g_4$ owing to the two dimensionality
based on the $g$-ology language.
The present study indicates that the cLC can emerge in metals near the magnetic criticality with geometrical frustration
\end{abstract}

\address{
Department of Physics, Nagoya University,
Furo-cho, Nagoya 464-8602, Japan. 
}
 
\sloppy

\maketitle

Various exotic symmetry-breaking phenomena
are recent central issues in strongly correlated metals.
For instance, violation of rotational symmetry,
so called the nematic order,
has been intensively studied in Fe-based 
\cite{Fernandes,Fernandes2,Chubukov-rev,Chubukov-AL,Onari-SCVC,Onari-FeSe,Yamakawa-FeSe}
and cuprate 
\cite{Schultz,Sachdev,Sachdev2,DHLee-PNAS,Kivelson-NJP,Yamakawa-Cu,Kawaguchi-Cu,Tsuchiizu-Cu1,Tsuchiizu-Cu2,Chubukov-Cu}
superconductors in addition to heavy fermion compounds
\cite{Ikeda-URu2Si2,Tazai-CeB6}.
Many kinds of even-parity and time-reversal invariant
unconventional order, such as the orbital order
\cite{Onari-SCVC,Onari-FeSe,Yamakawa-FeSe},
the $d$-wave bond order 
\cite{Schultz,Sachdev,Sachdev2,DHLee-PNAS,Kivelson-NJP,Yamakawa-Cu,Kawaguchi-Cu,Tsuchiizu-Cu1,Tsuchiizu-Cu2}
and the spin-nematic order
\cite{Fernandes,Fernandes2,Chubukov-rev,Chubukov-AL},
have been proposed as the candidate for the nematic order.
(Bond order is the symmetry breaking in 
correlated hopping integrals.)
Although the microscopic mechanism of the nematicity
is still under debate,
it is believed that many-body effects 
beyond the mean-field theory are significant
\cite{Fernandes,Fernandes2,Chubukov-rev,Chubukov-AL,Onari-SCVC,Onari-FeSe,Yamakawa-FeSe,Schultz,Sachdev,Sachdev2,DHLee-PNAS,Kivelson-NJP,Yamakawa-Cu,Kawaguchi-Cu,Tsuchiizu-Cu1,Tsuchiizu-Cu2,Chubukov-Cu,Ikeda-URu2Si2,Tazai-CeB6}.

When the unconventional order violates the 
parity and/or time-reversal symmetries,
more exotic phenomena emerge.
For example, the parity-violating bond order
induces the spontaneous spin current
\cite{Nersesyan,Kontani-sLC}.
Also, the time-reversal violating order
causes the static charge current,
which accompanies the internal magnetic field that is 
measurable experimentally.
Various charge-loop-currents (cLCs),
like the intra-unit cell cLC
\cite{Varma1,Varma2}
and antiferro-cLC 
\cite{Affleck,Bulut-cLC,Yokoyama,TKLee,Ogata,FCZhang,2leg-gia},
have been discussed.
In square lattice models, the cLC due to nonzero spin chirality has been studied based on SU(2) gauge theory \cite{conical,orbcurrent}. In addition to that, the generalized Hubbard ladder system has been studied by fRG and bosonization \cite{gene-ladder1,gene-ladder2}.

Recently, a number of experimental evidences for the cLC order have been reported.
For instance, in quasi 1D two-leg ladder cuprates,  
the polarized neutron diffraction (PND) reveal the broken time-reversal symmetry
\cite{cLC-2leg} and conclude that the cLC appears.
The cLCs are also reported in cuprates \cite{TRSB-neutron1,TRSB-neutron2} and iridates \cite{TRSB-iridate} by PND studies, and their
existences are supported by the optical second harmonic generation (SHG) 
\cite{SHG-cuprate,SHG-iridate}, Kerr effect \cite{Kerr-cuprate} and magnetic torque \cite{torque-iridate} measurements.

These observations indicate the existence of a universal mechanism of the cLC that is closely related to the magnetic criticality.
However, its microscopic origin is still unknown.
Based on a simple Hubbard model with on-site Coulomb interaction $U$,
mean-field theories fail to explain the cLC.
Therefore,
off-site Coulomb and Heisenberg interactions
have been analyzed
\cite{Nersesyan,Bulut-cLC}.
However, 
off-site bare interaction is much smaller than $U$ in usual metals.
Then, we encounter essential questions;
What is the minimum model to understand the cLC ?
What is the relation between cLC and magnetic criticality?

\begin{figure}[htb]
\includegraphics[width=.98\linewidth]{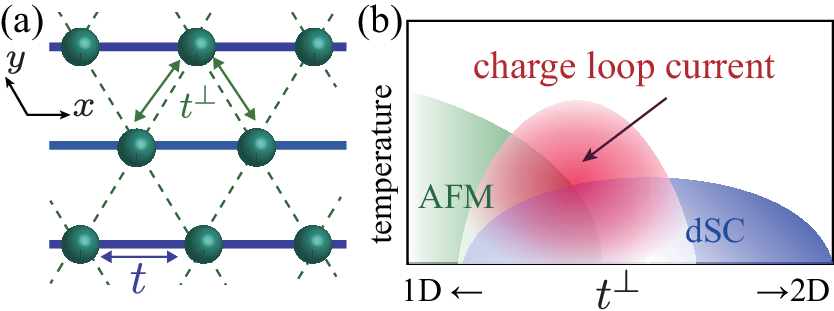}
\caption{
(a) Lattice structure with intra- ($t$) and inter- ($t^{\perp}$) chain hoppings.
(b) Schematic phase diagram. The cLC phase appears between AFM and $d$SC phase.
}
\label{fig:model}
\end{figure}

In this paper,
we propose a novel spin-fluctuation-driven cLC
mechanism based on the functional renormalization group 
(fRG) theory \cite{Metzner,Metzner2,Metzner3,Honnerkamp,Honnerkamp2,Honnerkamp3,Tsuchiizu-RG,Tsuchiizu2015,Tazai-RG,fRG-lect}.
Here, we optimize the form factor, which characterizes the essence of
the unconventional order,  unbiasedly
based on the Lagrange multipliers method.
By virtue of this method, the ferro-cLC order is obtained without bias
in a simple frustrated chain Hubbard model given in Fig. \ref{fig:model}(a).
We discover that the cLC appears 
between the antiferromagnetic (AFM) phase and
$d$-wave superconducting ($d$SC) phase
as schematically shown in Fig. \ref{fig:model}(b).
The present theory
indicates that cLC can emerge in strongly correlated electron systems with geometrical frustration.

The dimensional crossover in the coupled chain model
has been studied intensively for years
\cite{Emery,Bourbonnais,Kishine1,Kishine4,Suzumura,Suzumura2}.
For $T\gg t^{\perp}$, each Hubbard chain is essentially independent
because the thermal de Broglie wavelength is extremely short.
For $T\ll t^{\perp}$, inter-chain coherence is established,
and therefore quasi two-dimensional Fermi liquid (FL) state 
with finite quasi-particle weight is realized.
In the latter case, one-loop fRG method is very useful since
the incommensurate nesting vector of the Fermi surface (FS)
is accurately incorporated into the theory.
In the $g$-ology language
\cite{Emery,Bourbonnais,Kishine1,Kishine4,Suzumura,Suzumura2,Solyom}, 
the cLC order in the present theory is caused by the strong renormalization of the 
forward scatterings $g_2$ and $g_4$ owing to the two dimensionality.

Here, we study quasi-one-dimensional (q1D) electron systems described by
$\hat{H}=\hat{H}_{0}+\hat{H}'$.
The kinetic term is $\hat{H}_{0}=\sum_{\k\sigma}\epsilon_{\k}c^{\dagger}_{\k\sigma}c_{\k\sigma}$
where $c^{\dagger}$ is a creation 
operator for the electron with the momentum $\k$ and spin $\sigma$.
The energy dispersion is simply given by $\epsilon_{\k}=-2t\cos k_{x}-2t^{\perp}
\{ \cos k_{y}+\cos(k_{x}+k_{y})\}-\mu$ with $t=1$ and the chemical potential
$\mu$.
The inter-chain hopping  $t^{\perp} (\ll 1)$ controls the dimensionality;
$t^{\perp}\rightarrow 0$ corresponds to complete 1D system.
We also introduce on-site Coulomb interaction
$\hat{H}'=\sum_{i}U
n_{i\uparrow}n_{i\downarrow}$ where $i$ is the site index.

\begin{figure}[htb]
\includegraphics[width=.99\linewidth]{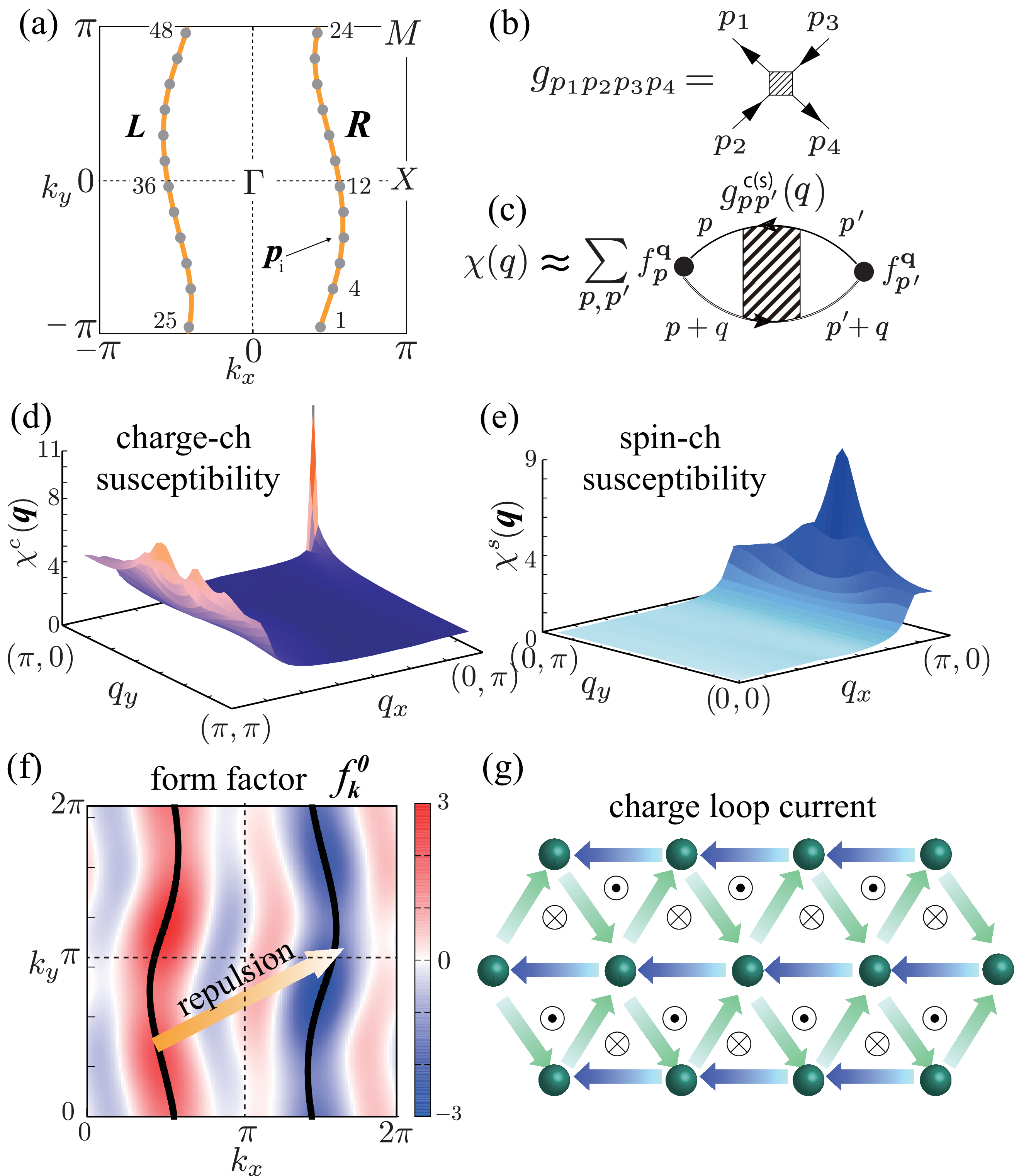}
\caption{
(a) FS with center positions of the patches.
(b) 4-point vertex function $\hat{g}$.
(c) Susceptibility with the form factor $f^{\q}_{\p}$. 
(d) Development of $\chi^{c}(\q)$ at $\q=\bm{0}$, which leads to the ferro-cLC.
(e) Spin susceptibility with the peak at $\q=(\pi,\pi/2)$.
(f) Obtained charge-channel form factor 
$f_{\k}^{\q=\bm{0}} (\propto \sin k_{x}+b\sin 3k_{x})$.
(g) Schematic picture of the cLC.  }
\label{fig:chi}
\end{figure}

\begin{figure}[htb]
\includegraphics[width=.99\linewidth]{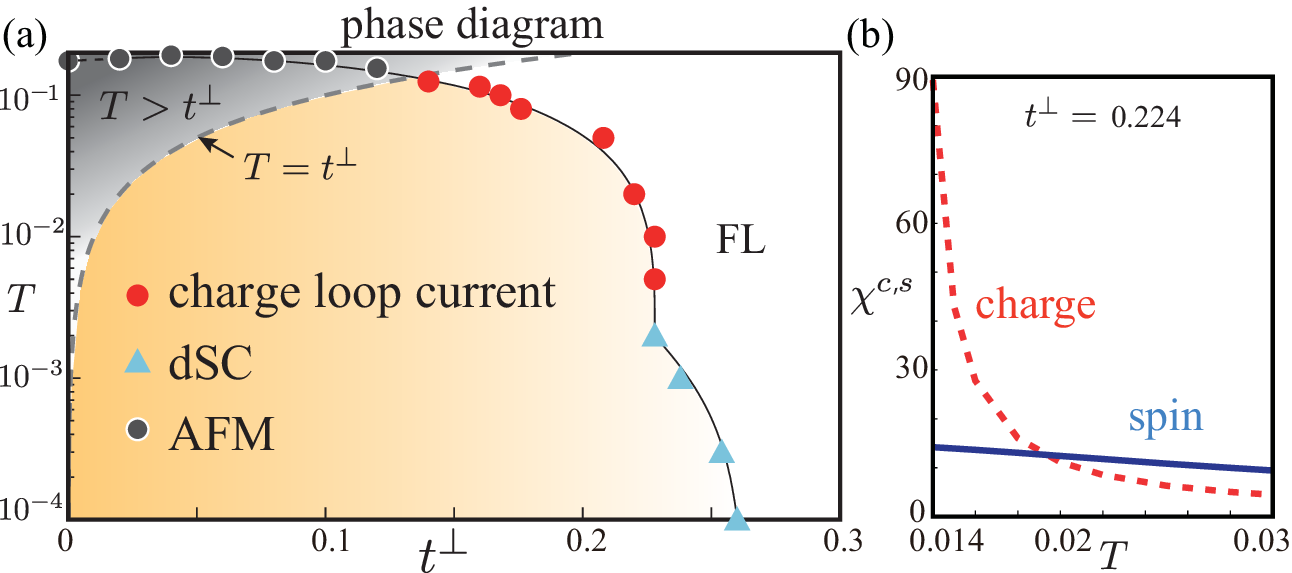}
\caption{ (a) Obtained phase diagram.
The cLC appears in the FL regime.
(b) Temperature dependence of the susceptibilities for the 
charge-ch (dotted line) and spin-ch (solid line).}
\label{fig:phase}
\end{figure}

Now, we perform the fRG method to derive effective low-energy interaction.
In the present numerical study, we divide each
left-Brillouin zone (BZ) and right-BZ into $24$ patches.
The center points of the patches $\bm{p}_i$ are 
shown in Fig. \ref{fig:chi}(a) and the Supplemental Materials A (SM.A) \cite{SM}.
Here, we introduce logarithmic energy scaling parameter $\Lambda_{l}=\Lambda_0 e^{-l}
 (0\leq l \leq l_c )$ for $\Lambda_0=3$, which is slightly larger than
$\max_{\k}{|\epsilon_{\k}|}\simeq 2.8$.
In the following numerical study, 
we consider the half-filling case and put $l_c=8.7 $ ($\Lambda_{l_c}=T/100$) and $U=2.01$ in the unit $t=1$.
Also, we fix $(t^{\perp},T)=(0.2,0.05)$  except for the phase diagram in Fig. \ref{fig:phase}.
During the fRG analysis, the low-energy effective interaction changes
with the cut off $\Lambda_{l}$. 
It is represented on the patches as
\begin{eqnarray}
\hat{H}'_{\rm{eff}}=\frac{1}{4}{\sum}_{\{p_{i}\} }g_{p_1 p_2 p_3 p_4} c^{\dagger}_{p_1}c_{p_2}c_{p_3}c_{p_4}^{\dagger} ,
\end{eqnarray} 
where $\hat{g}$ is antisymmetric 4-point vertex function with
patch $\p_{i}$ and spin index where $p_{i}\equiv (\p_{i},\sigma_{i})$.
$\hat{g}$ is defined in Fig. \ref{fig:chi}(b), and its
initial condition is $g_{p_1p_2p_3p_4}=U\delta_{\p_1+\p_4,\p_2+\p_3}(\delta_{\sigma_1, \sigma_3}\delta_{\sigma_2, \sigma_4}-\delta_{\sigma_1, \sigma_2}\delta_{\sigma_3, \sigma_4})$.
Then, $\hat{g}$ is calculated by solving the 1-loop RG equation,
\begin{eqnarray}\hspace{-5pt}
\frac{d}{d\Lambda_l}
g_{p_1p_2p_3p_4}&\hspace{-10pt}=\hspace{-10pt}&
\sum_{pp'}
\Bigl[ \frac{1}{2}\frac{dW_{p,p'}^{-}}{d\Lambda_l}
g_{p_1pp'p_4}g_{pp_2p_3 p'} \label{RGeq4} \\
+&&\hspace{-10pt} \frac{d W_{p,p'}^{+}}{d\Lambda_l}
\Bigl( g_{p_1p_3pp'} g_{pp'p_2p_4} 
-g_{p_1p_2pp'} g_{pp'p_3p_4} \Bigr) 
\Bigr], \nonumber 
\end{eqnarray} 
where 
$W_{p,p'}^{\pm}= T\sum_{\bm{k} \bm{k}' n}G_{\bm{\k} n}G_{\bm{\k}' \pm n}
\Omega_{\bm{p}}(\bm{k})\Omega_{\bm{p}'}(\bm{k}')$.
Here,   
$G_{\bm{k} n}\equiv (i\epsilon_n-\epsilon_{\bm{k}})^{-1}
\theta(\Lambda_{l}-|\epsilon_{\k}|)$,
and $\Omega_{\bm{p}}(\bm{k})= 1 \ (0)$ only if the momentum $\bm{k}$ 
is inside (outside) the $\bm{p}$-patch.
Here, $\epsilon_n$ is fermion Matsubara frequency.
The 1st term of r.h.s of the Eq. (\ref{RGeq4}) is the
particle-particle loop (=Cooper-channel (ch)),
and the 2nd and 3rd terms are the particle-hole loops (=Peierls-ch).
Their diagrammatic expressions are in SM.A \cite{SM}.

Here, we calculate the particle-hole susceptibilities, which are
essentially given by the 4-point vertex function in Fig. \ref{fig:chi}(c).
The static charge (spin)-ch susceptibilities with the form factor 
$f_{\k}^{\q}$ is defined by
\begin{eqnarray}
\chi^{c(s)}(\q)&=&\int^{T^{-1}}_{0} d\tau \frac{1}{2}\left\langle A^{c(s)}({\bm q},\tau)A^{c(s)}({\bm -\q},0)\right\rangle,   \label{eqn:chisc} \nonumber \\
A^{c(s)}({\bm q},\tau)&\equiv& \sum_{\bm k \sigma \sigma'}
\sigma^{0(z)}_{\sigma \sigma'} f_{\k}^{\q} c^{\dagger}_{{\k+\q}  \sigma}(\tau)
c_{{\k} \sigma'}(\tau), \label{eq:chi2} \end{eqnarray}
where  $\tau$ is imaginary-time.
$\hat{\sigma}^{0}$ is identity matrix and $\hat{\sigma}^{z}$ is 
the Pauli matrix. 
Here, we optimize the form factor $f_{\k}^{\q}$
unbiasedly to maximize $\chi^{c}(\q)$ at each $\q$-point 
using the Lagrange multipliers method; see the SM.A \cite{SM}.

The form factor corresponds to the 
modulation of the correlated hopping integral
from $j$ to $i$-site, $\delta t_{ij} (i \neq j)$.
It is given as
$\delta t_{ij}= \Delta t \sum_{\k} f_{\k}^{\q=\bm{0}} e^{i\k (\bm{r}_{i}-\bm{r}_{j})}$ for the
uniform modulation. 
Here, both the bond order
($\delta t_{ij}=\delta t_{ji}$) and the cLC
($\delta t_{ij}=-\delta t_{ji}$) are described.
Due to the Hermite condition, $\delta t_{ij}$ for the cLC order is pure imaginary.
In Figs. \ref{fig:chi}(d) and (e), we plot the $\q$-dependence of the charge- and spin-ch susceptibilities, respectively.
The strong charge-ch fluctuations develop at $\q=\bm{0}$,
while the spin fluctuations remain small even
at the peak $\q=(\pi,\pi/2)\equiv \bm{Q}_{\rm{AFM}}$.
Figure \ref{fig:chi}(f) shows the
charge-ch form factor
at $\q=\bm{0}$.
For a fixed $k_y$, the
relation $f^{\bm{0}}_{k_x}\simeq -f^{\bm{0}}_{-k_x} (\propto \sin k_x+b\sin 3k_x )$ holds.
Then, the real-space order parameter is 
$\delta t_{ij}=-\delta t_{ji}$ that leads to
 the emergence of ferro-cLC order. 
The third-nearest-intra-chain form factor derived from the present fRG
is significant for realizing the cLC \cite{SM}.
In Fig. \ref{fig:chi}(g), we show the schematic picture of the cLC, which is
a magnetic-octupole-toroidal order. The detailed explanation of the numerical
results are shown in Fig. S4 in SM.B \cite{SM}. 

In Fig. \ref{fig:phase}(a), obtained phase diagram in the $T$-$t^{\perp}$ space is plotted.
We reveal that the cLC phase appears around $t^{\perp}\simeq 0.2$
as an intertwined order between AFM and $d$SC states.
Note that the dark shaded area is 1D Mott insulating phase
that is beyond the scope of the present study \cite{Kishine1,Kishine4}.
In addition, Fig. \ref{fig:phase}(b) shows the $T$-dependence of the
$\chi^{c}(\bm{0})$ and $\chi^{s}(\bm{Q}_{\rm{AFM}})$.
$\chi^{c}(\bm{0})$ drastically develops at low temperatures.
The transition temperatures in Fig. \ref{fig:phase}(a) are
determined under the condition that the largest susceptibility
(spin, charge, $d$SC) exceeds $\chi_{\rm{max}}=30$,
while the phase diagram is insensitive to $\chi_{\rm{max}} \ (\gg 10)$,
as recognized in Fig. \ref{fig:phase}(b).
As a result, the cLC phase is stabilized in the FL region around $t_{\perp} \gg T$.

\begin{figure}[htb]
\includegraphics[width=.995\linewidth]{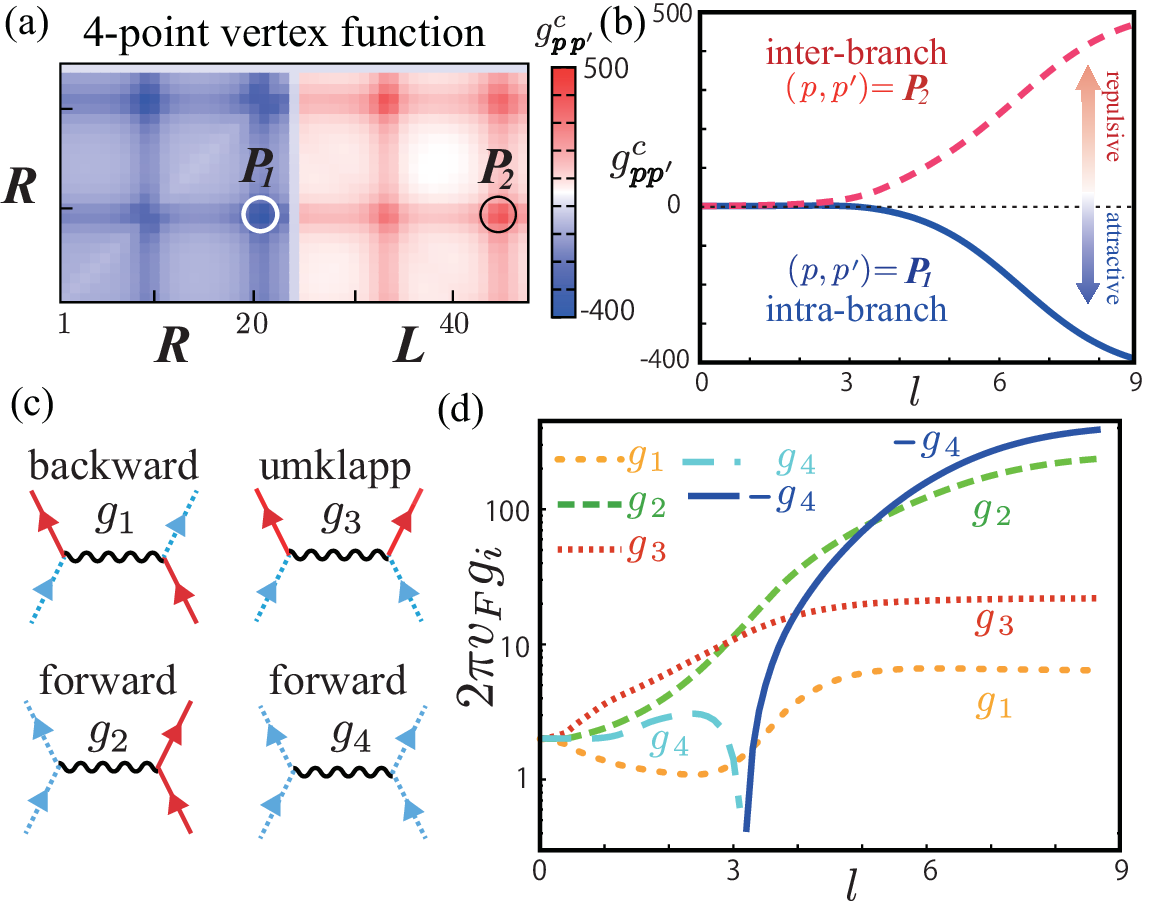}
\caption{
(a) Patch-dependence of four-point vertex at $\q=\bm{0}$.
The obtained relation $g^{c}_{\bm{R R}}(\bm{0})\approx 
-g^{c}_{\bm{L R}}(\bm{0})$ gives the ferro-cLC.
(b) Flow of $g^{c}_{\p \p'}(\bm{0})$.
$g^{c}_{\bm{R R (LR)}}(\bm{0}) $ takes large negative (positive) value as the cutoff
energy decreases.
(c) Definition of the $g_{i}$ in the $g$-ology theory.
Dotted (solid) line gives electron on the right (left) branch.
(d) Obtained flow of $g_i$.
}
\label{fig:golog}
\end{figure}

To understand the origin of the cLC, we analyze
the charge (spin)-ch 4-point vertex function defined by 
\begin{eqnarray}
g^{c(s)}_{\p \p'}(\q)\equiv g_{\p \uparrow \p+\q \uparrow \p' \uparrow \p'+\q \uparrow}+(-)g_{\p \uparrow \p+\q \uparrow \p' \downarrow  \p'+\q \downarrow}.\end{eqnarray}
In Fig. \ref{fig:golog}(a), we plot the patch-dependence of the charge-ch 4-point vertex
$g^{c}_{\bm{p p'}}(\bm{0})$.
The relation $g^{c}_{\bm{R R'}}(\bm{0}) \approx -g^{c}_{\bm{L R}}(\bm{0})$ holds,
where $\bm{R}=1\sim 24\ (\bm{L}=25\sim 48)$ 
is patch index in the right (left) branch.
We also plot the flow ($l$-dependence) of the 4-point vertex in Fig. \ref{fig:golog}(b).
$g_{\bm{R R'}}^{c}(\bm{0})$ comes to be large negative value, while
$g_{\bm{L R}}^{c}(\bm{0})$ takes large positive value.

In order to explain why odd-parity form factor is obtained,
we introduce $\bar{g}^{c(s)}_{RR}(\q)$, $\bar{g}^{c(s)}_{LR}(\q)$,
$\bar{f}^{\bm{0}}_{{R}}$, $\bar{f}^{\bm{0}}_{{L}}$ as their maximum values 
in the patch space.
In this case, the charge-ch susceptibility is
\begin{eqnarray}
\chi^{c}(\bm{0})\propto
- (\bar{f}^{\bm{0}}_{{R}})^2 \bar{g}^{c}_{{R R}}(\bm{0})-\bar{f}^{\bm{0}}_{{L}} \bar{f}^{\bm{0}}_{{R}} \bar{g}^{c}_{{L R}}(\bm{0})
\end{eqnarray}
as shown in Fig. \ref{fig:chi}(c).
Since $\bar{g}^{c}_{{R R}}(\bm{0})$ is negative and 
$\bar{g}^{c}_{{L R}}(\bm{0})$ is positive,
the relation $\bar{f}^{\bm{0}}_{{R}}=-\bar{f}^{\bm{0}}_{{L}}$
is required to maximize the susceptibility.
In conclusion, the odd-parity cLC appears due the
sign reversal between $\bar{g}^{c}_{{R R}}(\bm{0})$ and $\bar{g}^{c}_{{L R}}(\bm{0})$
in the FL region.
As for the spin-ch susceptibilities,
both $g_{{R R}}^{s}(\bm{Q}_{\rm{AFM}})$ and $g_{{L R}}^{s}(\bm{Q}_{\rm{AFM}})$ are negative, and therefore the spin-ch form factor,
does not have any sign reversal on the FS 
as shown in Fig. S3 in SM.A \cite{SM}.
Thus, ordinal AFM phase is realized in the 1D regime.

Here, we discuss the present result in terms of the 1D $g$-ology theory
\cite{Bourbonnais}, in which 
the 4-point vertex function is classified into 4-types;
backward ($g_1$), forward ($g_2,g_4$) and umklapp ($g_3$) scatterings as 
defined in Fig. \ref{fig:golog}(c).
As an approximation, there is  one-to-one correspondence 
between  $\bar{g}_{p p'}^{c(s)}(\q)$ and $g_{i} (i=1-4) $ as
\begin{eqnarray}
&\!\!\!\!\!\! \bar{g}^{c}_{RR}(\bm{0})\approx 2\pi v_F g_4, \hspace{3pt}
&\!\!\! \bar{g}^{c}_{LR}(\bm{0})\approx 2\pi v_F(2g_2 -g_1), \nonumber \\
&\!\!\!\!\!\! \bar{g}^{s}_{RR}(\bm{Q}_{\rm{AFM}})\approx -2\pi v_F g_2, \hspace{3pt}
&\!\!\! \bar{g}^{s}_{LR}(\bm{Q}_{\rm{AFM}})\approx  -2\pi v_F g_3 \label{eq:golo1},
\end{eqnarray}
where 
$v_{F}$ is the Fermi velocity.

Based on the Eq.(\ref{eq:golo1}), we plot the flow of $g_i$ in Fig. \ref{fig:golog}(d).
We find that $g_4$ ($g_2$) has large negative (positive) value as the $l$ increases.
The present result is understood by using the knowledge of the $g$-ology theory
as we discuss in SM.D \cite{SM}: 
At half filling, $g_2$ is relevant due to the
Peierls-ch scattering \cite{Bourbonnais}.
In the present q1D model, the frustrated hopping $t^{\perp}$ 
violate the perfect nesting condition, and therefore
$g_2$ (or AFM fluctuation) is relatively suppressed  at $\Lambda_l < t^{\perp}$
compared with pure 1D systems \cite{Emery,Bourbonnais,Kishine1,Kishine4,Suzumura,Suzumura2}.
On the other hand, surprisingly, $g_4$ takes large negative values due to the
Landau-ch scattering that is important at low energies ($\Lambda_l < T$). 
As a result, 1D AFM instability is suppressed by $t^{\perp}$, 
and the cLC due to the Landau-ch instead appears.
(Importance of the $g_4$ on $\chi^s(\bm{0})$
was discussed in Ref. \cite{fuseya_g4}.)
Thus, the geometrical frustration is essential for 
realizing the cLC order.
\begin{figure}[htb]
\includegraphics[width=.99\linewidth]{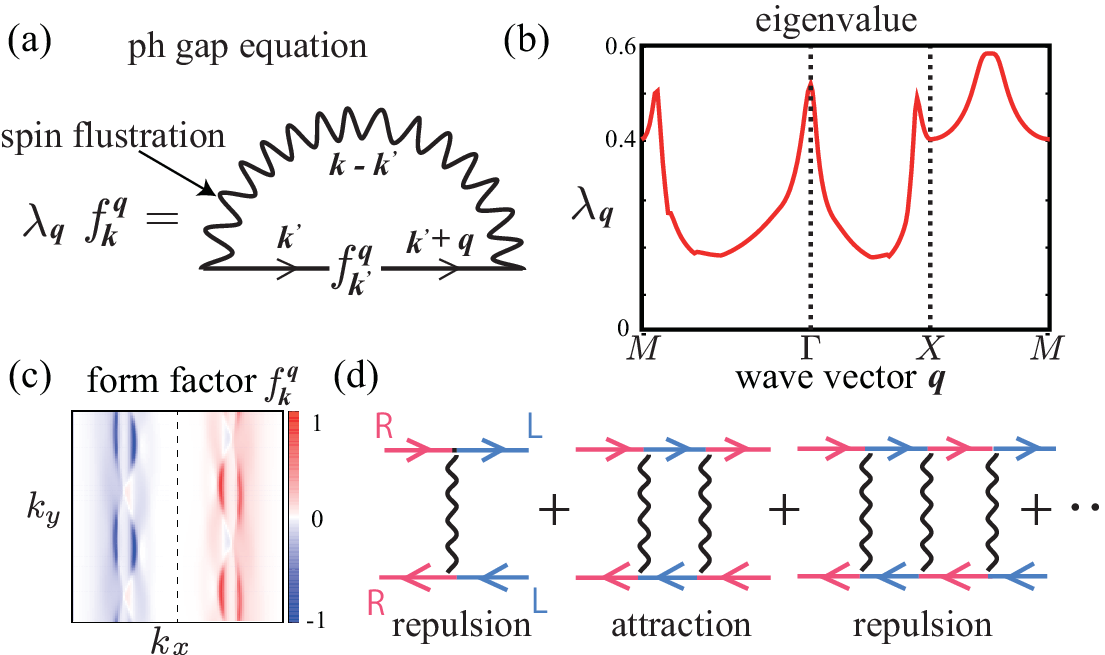}
\caption{
(a) The ph gap equation. Wavy line is spin fluctuations given by 
the RPA.
(b) Obtained eigenvalue of the gap equation at $(T,U)=(0.03,1.65)$. 
The peak around $\Gamma$-point corresponds to the cLC.
(c) Obtained charge-ch form factor at $\q=\bm{0}$, which is essentially the same as
the fRG results. (d) Dominant contribution for stabilizing the
cLC order.}
\label{fig:cdw}
\end{figure}

Also, the cLC is naturally understood by
the spin-fluctuation-driven mechanism based on 2D FL concept
\cite{Kontani-rev,Moriya,Yamada}.
To show this, we solve the ''particle-hole (ph) gap equation'' for
charge-ch form factor $f^{\q}_{\k}$ ;
\begin{eqnarray}
\lambda_{\q}  f^{\q}_{\k}=\sum_{\k'} f^{\q}_{\k'} L(\k',\q) \left(-\frac{3}{2}V^{s}_{\k-\k'} -\frac{1}{2}V^{c}_{\k-\k'} \right),\label{eq:cdw}
\end{eqnarray} where $\lambda_{\q}$ is the eigenvalue.
Here, we define $V^{c(s)}_{\q}\equiv -(+)U+U^2 \chi^{c(s)}(\q)$ in the random-phase approximation (RPA),
and $L(\k,\q)\equiv (n_{\k-\frac{\q}{2}}-n_{\k+\frac{\q}{2}})/(\epsilon_{\k+\frac{\q}{2}}- \epsilon_{\k-\frac{\q}{2}}) >0$ with Fermi distribution function $n_{\k}$.
The diagrammatic expression of the ph gap equation in Fig. \ref{fig:cdw}(a) is 
given by the 1st order spin fluctuation exchange term ($=$MT-type process).
Figure \ref{fig:cdw}(b) shows the largest eigenvalue
$\lambda_{\q}$ for general $\q$.
The 2nd largest peak at $\Gamma$-point corresponds to the cLC
since the obtained odd-parity form factor in Fig. \ref{fig:cdw}(c)
is essentially the same as the results by the fRG. 
(Obviously, the fRG method is superior to RPA in that loop cancellation in 1D system
 is taken into account.)
Figure \ref{fig:cdw}(d) shows the scattering processes
generated by solving the ph gap equation.
The even (odd)-order processes with respect to $\chi^s(\bm{Q}_{\rm AFM})$
work as inter-branch repulsion (intra-branch attraction),
and it corresponds to $\bar{g}^{c}_{{RR}}<0 \ (\bar{g}^{c}_{{LR}}>0 )$ in Fig.\ref{fig:golog} (a).
Thus, the cLC is naturally explained in terms of the FL concept,
and this mechanism is found to be similar to that
for the $d$SC near the AFM phase \cite{Kishine1,Kishine4,Suzumura,Suzumura2,Kino}.

Furthermore, 
we perform the fRG without Cooper-ch processes and confirm that 
the cLC is obtained even if we neglect the Cooper-ch as shown in Fig. S5
in SM.C \cite{SM}.
Thus, we conclude that the cLC emerges due to the spin-fluctuation-driven mechanism.
\begin{figure}[htb]
\includegraphics[width=.99\linewidth]{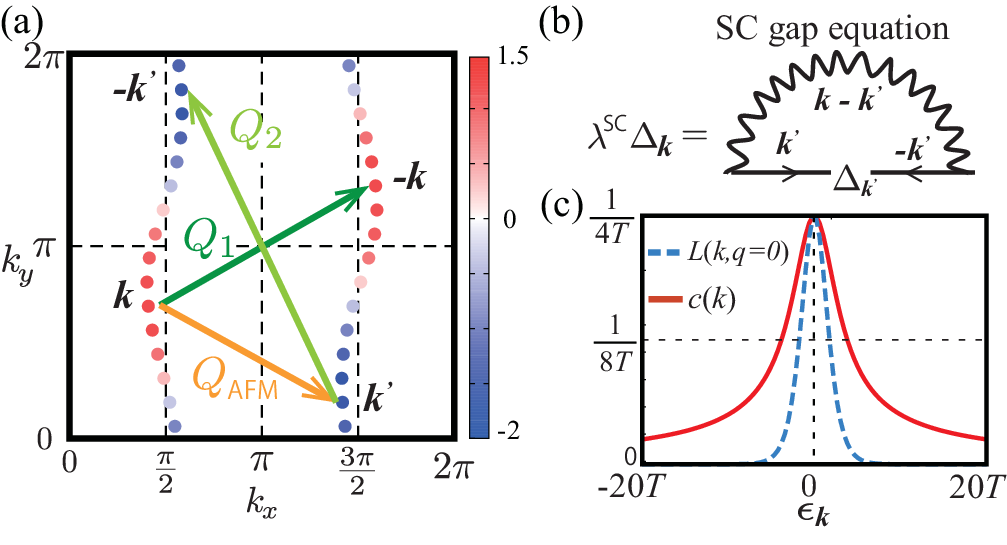}
\caption{
(a) $d$SC gap function obtained by the fRG.
(b) SC gap equation.
(c) $\epsilon_{\k}$-dependence of $L(\k,\q=0)$ and $C(\k)$.}
\label{fig:sc}
\end{figure}

Next, we discuss the $d$SC phase.
Figure \ref{fig:sc}(a) shows the optimized SC gap
given by the fRG \cite{Tsuchiizu-Cu1}.
This $d$SC gap is well understood 
in terms of the singlet SC gap equation with the MT-process \cite{Kino}; 
\begin{eqnarray}
\lambda^{SC}  \Delta_{\k}= \sum_{\k'} \Delta_{\k'} C(\k') \left(-\frac{3}{2}V^{s}_{\k-\k'} +\frac{1}{2}V^{c}_{\k-\k'} \right),\label{eq:scgap}
\end{eqnarray}
where $C(\k)=(2n_{\k}-1)/(2\epsilon_{\k})>0$ and its diagrammatic expression 
is in Fig. \ref{fig:sc}(b). Since $\Delta_{\k}$ is even-parity, the
inter-branch repulsion by $V^{s}_{\bm{Q}_{\rm{AFM}}} $ induce the nodal $d$SC.

Furthermore, we discuss the reason why the cLC phase dominates over the SC phase.
As shown in Fig. \ref{fig:sc}(c), $C(\k)$ in Eq. (\ref{eq:scgap}) 
is always larger than $L(\k,\q=\bm{0})$ in Eq. (\ref{eq:cdw}) 
except at $\epsilon_{\k}=0$,
reflecting the logarithmic Cooper-ch singularity \cite{Bourbonnais}.
On the other hand, the Cooper instability is reduced by the intra-branch sign
reversal in the $d$SC gap.
By considering the dominant contribution of the gap 
function at $(\k,\k',-\k,-\k')$ in Fig. \ref{fig:sc} (a), 
the effective pairing interaction for the $d$-wave gap is
\begin{eqnarray} \hspace{-10pt}
V_{dSC} \hspace{-3pt}
\propto \{2\chi^s(\bm{Q}_{\rm AFM})-\chi^s (\bm{Q}_{1})-\chi^s (\bm{Q}_2) \}
\propto (t^{\perp}/t)^2. \label{Vdsc}
\end{eqnarray}
Thus, the $d$-wave Cooper instability
is suppressed by the factor $(t^{\perp}/t)^2 \ll 1$ due to the 1D nature.

If we consider off-site Coulomb interaction $V$ in addition to $U$, the cLC instability should be enhanced.
In fact,  the Fock term $-2V\cos(\k-\k') $ is added to $V^{c}_{\k-\k'}$ in Eq. (\ref{eq:cdw}),
and it gives the inter-branch repulsive and intra-branch attractive interactions \cite{Varma1,Varma2,Bulut-cLC,Nersesyan}.
Thus, both the spin fluctuation and finite off-site Coulomb $V$ 
will cooperatively stabilize the cLC phase.

In summary, we proposed the spin-fluctuation-driven cLC mechanism
 based on the fRG theory.
We derived the optimized form factor, which is the key essence of
the unconventional order, without any assumptions.
By virtue of this method, the ferro-cLC order is obtained without any bias
in a simple frustrated chain Hubbard model.
For the microscopic origin of the cLC,
strong renormalization of the 
forward scatterings ($g_2$,$g_4$) due to spin fluctuations plays an important role.
We stress that 
the cLC phase in the FL regions is replaced with the AFM phase
if we remove the frustration as shown in Fig. S9 in SM. E.
The role of geometrical frustration is to realize strong short-range spin fluctuations that mediate the cLC order. 
Thus, it will be useful to verify the theoretically predicted correlation between the cLC order and spin fluctuation strength in future experiments.

We are grateful to S. Onari and M. Tsuchiizu for useful discussions.
This work is supported by Grants-in-Aid for Scientific Research (KAKENHI) 
Research (No. JP20K22328, No. JP20K03858, No. JP19H05825, No. JP18H01175) 
from MEXT of Japan.

\clearpage
\newpage
\makeatletter
\renewcommand{\thefigure}{S\arabic{figure}}
\renewcommand{\theequation}{S\arabic{equation}}
\makeatother
\setcounter{figure}{0}
\setcounter{equation}{0}
\setcounter{page}{1}
\setcounter{section}{1}

\begin{widetext}
\begin{center}
{\bf 
[Supplementary Material] \\
Emergence of Charge Loop Current in Geometrically Frustrated Hubbard Model:\\
Functional Renormalization Group Study}%
\end{center}

\begin{center}Rina Tazai, Youichi Yamakawa and 
Hiroshi Kontani
\end{center}

\begin{center}
\textit{Department of Physics, Nagoya University, Nagoya 464-8602, Japan}
\end{center}
\end{widetext}
\subsection{A. fRG analysis, Optimization of form factor}
\begin{figure}[htb]
\includegraphics[width=.7\linewidth]{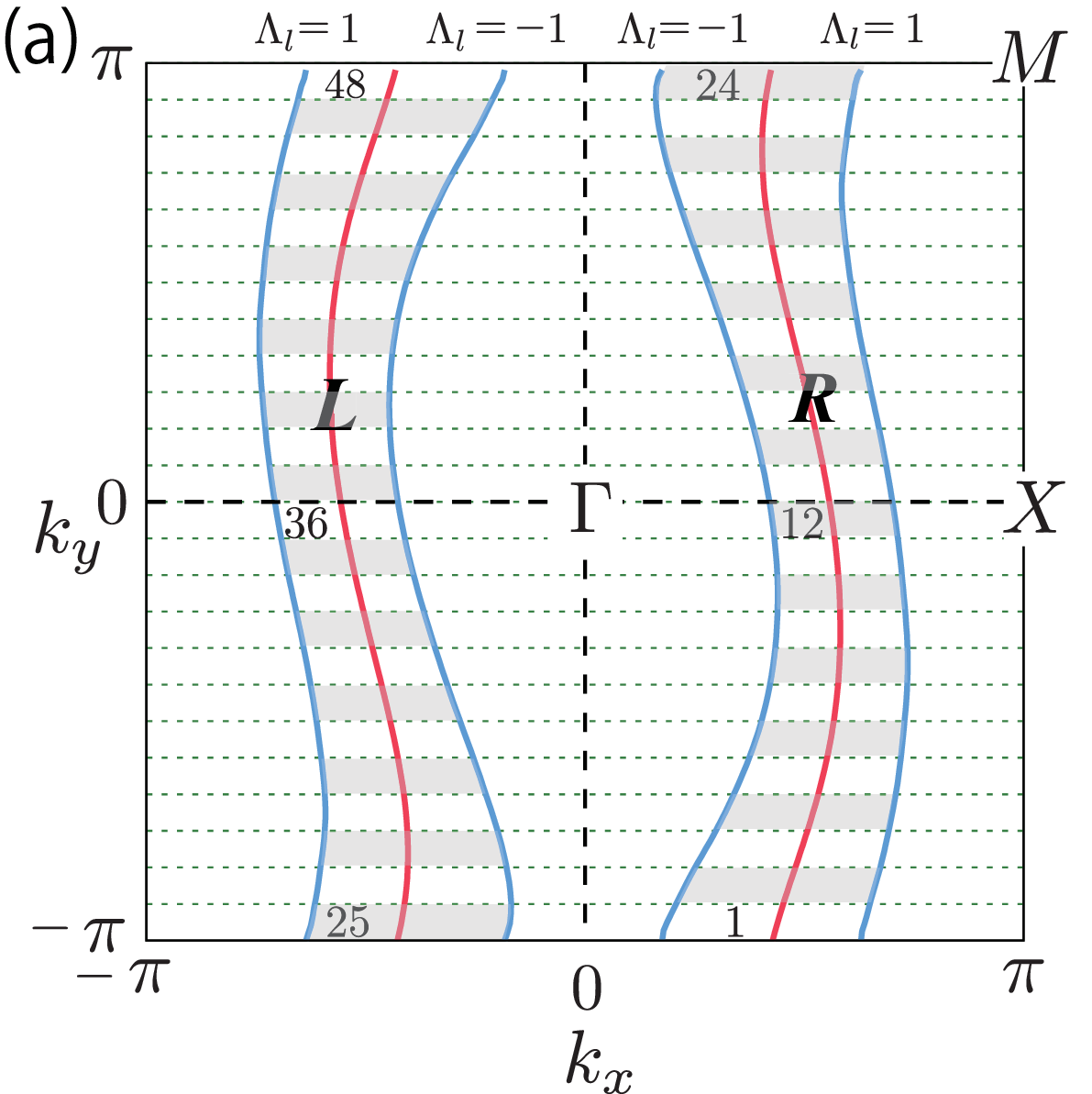}
\caption{
The 48-patches in the 1st BZ.
The red (blue) line shows the Fermi surface (boundary line at $\Lambda_l=\pm 1$).}
\label{fig:area}
\end{figure}

Here, we present the detailed explanation of the present fRG method.
In the framework of the patch-fRG,
the $\k$-space in the 1st BZ is divided into
$N_{p}$-patches.  In the present model, we set $N_{p}=48$,
and the boundary lines at $\Lambda_l=\pm 1$ are shown in Fig. \ref{fig:area}.
Each patch is rectangular along $k_x$-axis.

\begin{figure}[htb]
\includegraphics[width=.99\linewidth]{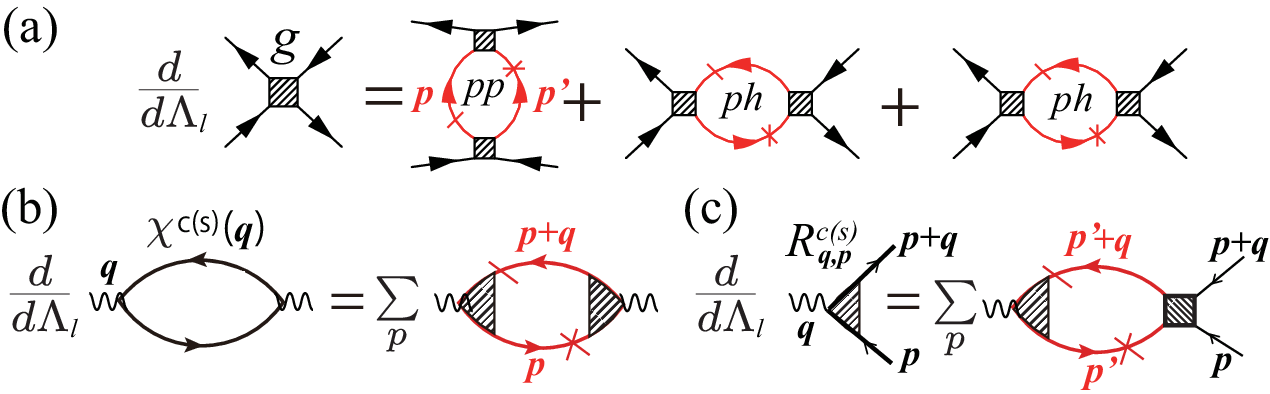}
\caption{
(a) RG equation for the 4-point vertex $\hat{g}$ and
that for the (b) charge (spin)-ch susceptibility $\chi^{c(s)}(\q)$
and (c) the 3-point vertex $R^{c(s)}_{\q,\p}$.}
\label{fig:fRGsap}
\end{figure}

Then, we solve the RG equation for the 4-point vertex function $\hat{g}$,
which is given in Eq. (\ref{RGeq4}) in the main text.
Its diagrammatic expression
is in Fig. \ref{fig:fRGsap}(a). The 1st (2nd and 3rd) term of the r.h.s 
correspond to the Cooper (Peierls)-ch scattering.
After that, the particle-hole susceptibilities
are obtained by solving the RG equation,
\begin{eqnarray}
\frac{d\chi^{c(s)}(\q)}{d\Lambda_{l}}
=T\sum_{\p}
\frac{d W^{+}_{\p,\p+\q}}{d\Lambda_{l}}
R^{c(s)}_{\q, \p} R^{c(s)}_{-\q, \p}, \label{eq:sapchi}
\end{eqnarray}
where $R^{c(s)}$ is the charge (spin)-ch 3-point vertex function and
obtained by
\begin{eqnarray}\frac{dR^{c(s)}_{\q,\p}}{d\Lambda_{l}}
=-T\sum_{\p'}
\frac{d W^{+}_{\p',\p'+\q}}{d\Lambda_{l}}
R^{c(s)}_{\q, \p'} g^{c(s)}_{\p'+\q, \p}(\q)\label{eq:sapR},
\end{eqnarray}
where $R^{c(s)}_{\q,\p}$ includes the form factor and
$W^{+}_{\p,\p+\q}$ is introduced in the main text.
The diagrammatic expression of Eqs. (\ref{eq:sapchi}) and (\ref{eq:sapR}) are
given in Figs. \ref{fig:fRGsap}(b) and (c), respectively.
Thus, the ph susceptibilities are essentially derived from
the 4-point vertex function as shown in Fig. \ref{fig:chi}(c) in the main text.

Here, we optimize the form factor $f_{\k}^{\q}$ so as to maximize the
ph susceptibility under the constraint $\sum_{\k} |f^{\q}_{\k}|^2=1$ at each $\q$-point. 
For this purpose, we introduce the Fourier expansion form of $f^{\q}_{\k}$ as
\begin{eqnarray}
f_{\k}^{\q}=\sum_{n,m=1}^{7}a_{n m}^{\q} h_{n}(k_x) 
h_m(k_y),\end{eqnarray}
where $\frac{h_{n}(k)}{\sqrt{2}}=\{ \frac{1}{\sqrt{2}}, \cos k, \cos 2k, \cos 3k,
\sin k, \sin 2k, \sin 3k \}$ for $n=1,2,3,4,5,6,7$, respectively.

Based on the Lagrange multipliers method,
the coefficient $a_{M}^{\q}$ is optimized under the condition
$\frac{1}{N}\sum_\k|f_{\k}^{\q}|^2=1$ 
by solving the following eigen equation,
\begin{eqnarray}
\sum_M \chi_{LM}^{c(s)}(\q) a^{\q}_{M} =\lambda a_{L}^{\q},
\end{eqnarray}
where the index $L \equiv (n,m)$ takes $1-49$.
The eigenvale $\lambda$ corresponds to the 
undetermined multiplier in the Lagrange multipliers method.
$\chi_{LM}^{c(s)}(\q)$ is the $(L,M)$ component of the 
charge-ch (spin-ch) susceptbility calculated by the fRG method:
\begin{eqnarray}
\chi^{c(s)}_{LM} (\q)&\equiv &\int^{\beta}_{0} d\tau \frac{1}{2}\left\langle B^{c(s)}_{L}({\bm q},\tau)B^{c(s)}_{M}({\bm -\q},0)\right\rangle,   \label{eqn:sapchi} \nonumber \\
B^{c(s)}_{nm}({\bm q},\tau)&\equiv & \sum_{\bm k \sigma \sigma'}
\sigma^{0(z)}_{\sigma \sigma'}h_n (k_x) h_m (k_y) c^{\dagger}_{{\k+\q}  \sigma}(\tau)
c_{{\k} \sigma'}(\tau). \label{eq:sapchi2} 
\nonumber 
\end{eqnarray}
Note that conventional charge (spin) susceptibility 
with $f_\k^\q=1$ is given by $L=M=(1,1)$.

\begin{figure}[htb]
\includegraphics[width=.5\linewidth]{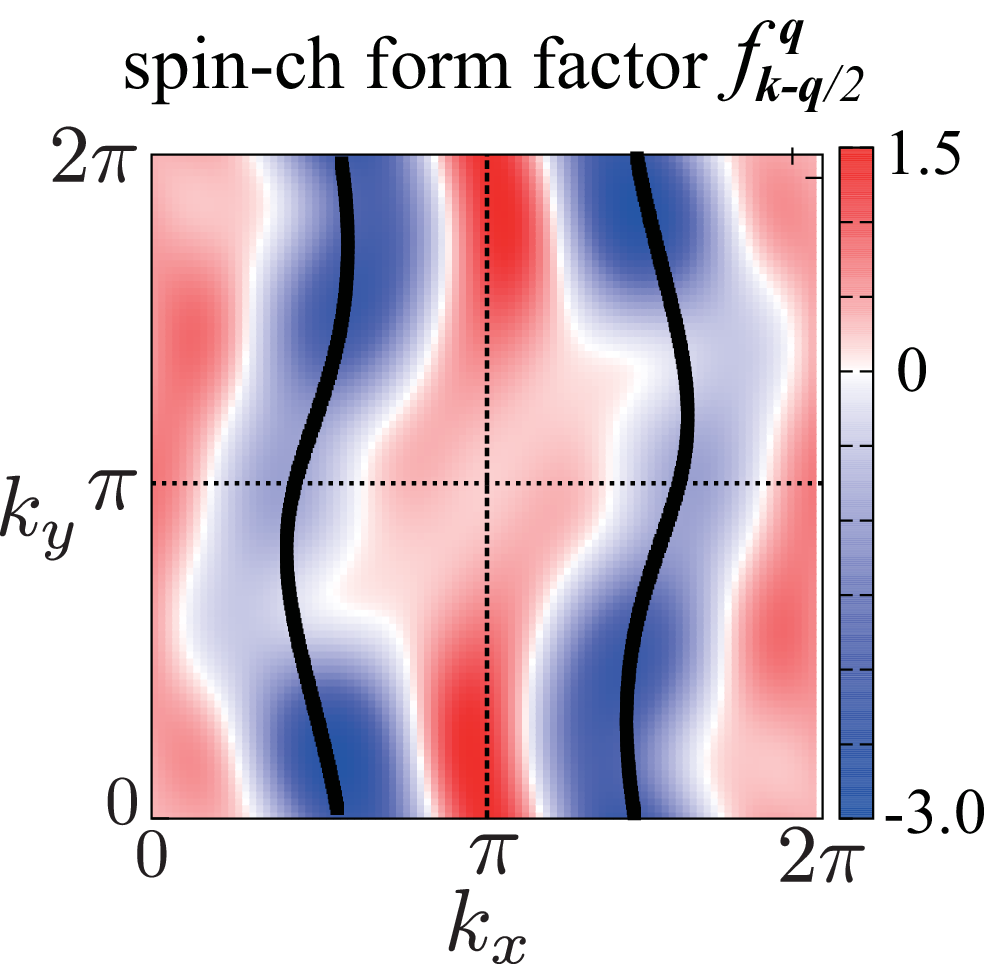}
\caption{
$\k$-dependence of the spin-ch form factor,
which corresponds to the conventional AFM fluctuation.}
\label{fig:sap_spin}
\end{figure}

In the main text, we show the optimized susceptibilities
and its form factor only for the charge-ch.
Here, we show the results for the spin-ch.
In Fig. \ref{fig:sap_spin}, we plot
spin-ch form factor $f^{\q}_{\k-\q/2}$ for $\q=\bm{Q}_{\rm{AFM}}$.
It does not have any sign reversal on the FS and
the conventional AFM fluctuations develop.
Thus, the obtained phase diagram in Fig. \ref{fig:phase}(a)
will be changed only slightly even if we consider the 
$\k$-dependent spin-ch form factor,
although the AFM phase will be slightly enhanced.
\subsection{B. Charge currents in real space}
Now, we show the detailed results of
the cLC from $0$-site to $i$-site ($\bm{r}=\bm{0}$),
which is defined by \cite{SKontani-sLC},
\begin{eqnarray}
j_i\equiv 2i e \left\{ (t_{i0} + \delta t_{i0})G (-\bm{r}_{i})- (t_{0i} + \delta t_{0i})G (\bm{r}_{i}) \right\},\end{eqnarray} where $-e$ is the charge of electron and
$\delta t_{i0}$ is obtained by Fourier transformation of charge-ch $f^{\q}_{\k}$
multiplied by the energy scale $\Delta t$.
Note that $\delta t_{i0}$ is pure imaginary
and $\delta t_{i0}=-\delta t_{0i}$ holds.
Here, equal-time Green function $G (\bm{r}_i)$ in real space is defined by
\begin{eqnarray}
G (\bm{r}_i)=T\sum_{n,\k} \frac{1}{i\epsilon_n-\epsilon_{\k}-\Delta t f_{\k}^{\bm{0}}}
e^{i\k \bm{r}_i}.
\end{eqnarray}

\begin{figure}[htb]
\includegraphics[width=.99\linewidth]{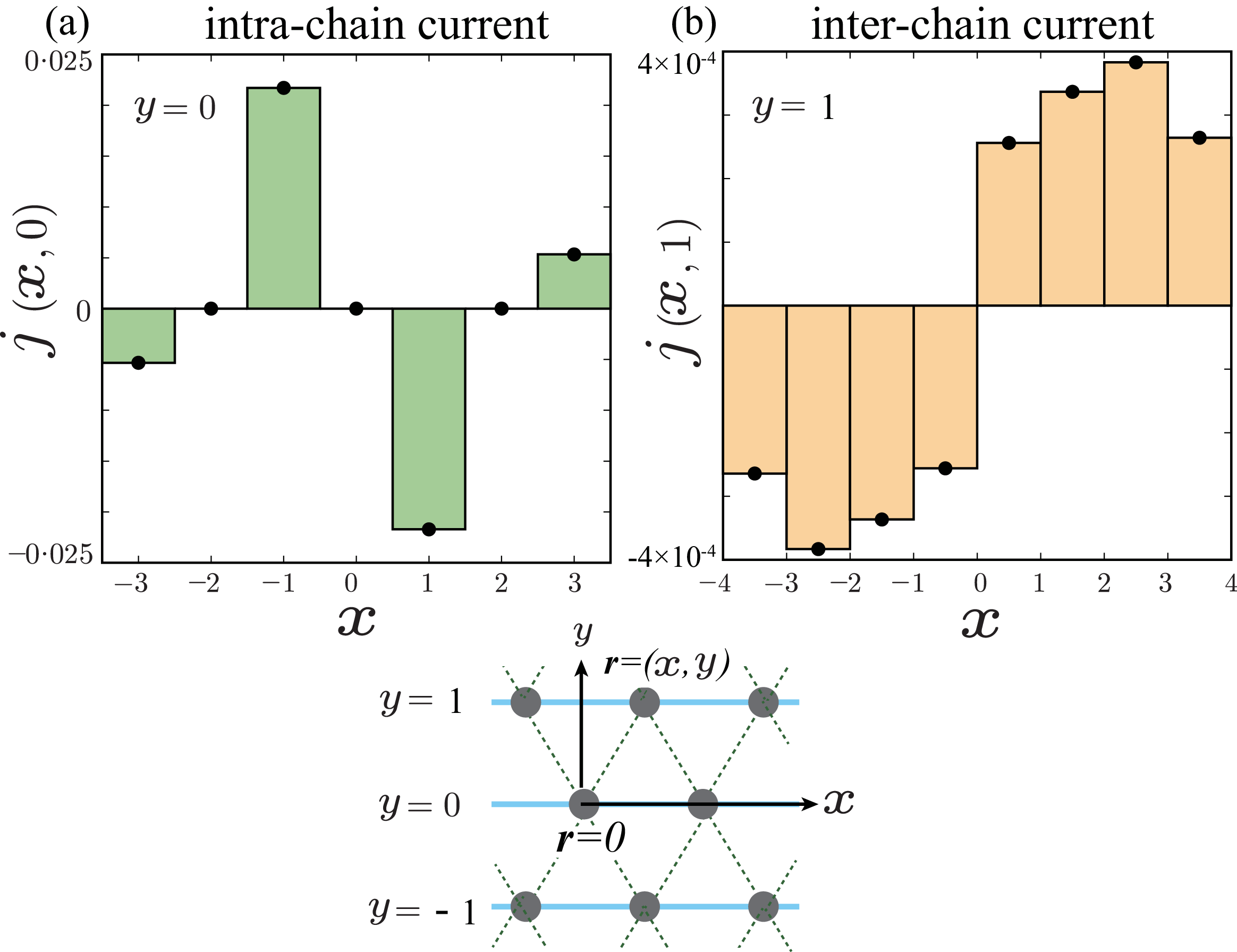}
\caption{
(a) Site-dependence of the intra-chain current $j_{x, y=0}$ and (b) that of 
inter-chain current $j_{x, y=1}$.}
\label{fig:sap_inter}
\end{figure}

Figures \ref{fig:sap_inter}(a) and (b) show the obtained values of the intra- and inter-chain components of the cLC, respectively. 
Here, we put $e=1$ and $\Delta t=0.05$.
The total intra-chain current is 
$j_{intra}=\sum_{x \geq 1} j(x,0) \times |x| \sim -7\times 10^{-3}$.
The schematic current pattern between the nearest sites is given in Fig. \ref{fig:chi}(g).
The cLC-induced magnetic field can be detected by 
experiments such as NMR.
We verified that the macroscopic current is zero due to
the cancellation between intra- and inter-chain current.
These results are consistent with the ''Bloch's theory''
that proved the absence of the macroscopic currents in the 
infinite periodic systems \cite{SBohm}.
\subsection{C. Analysis of the fRG without Cooper-ch}
\begin{figure}[htb]
\includegraphics[width=.55\linewidth]{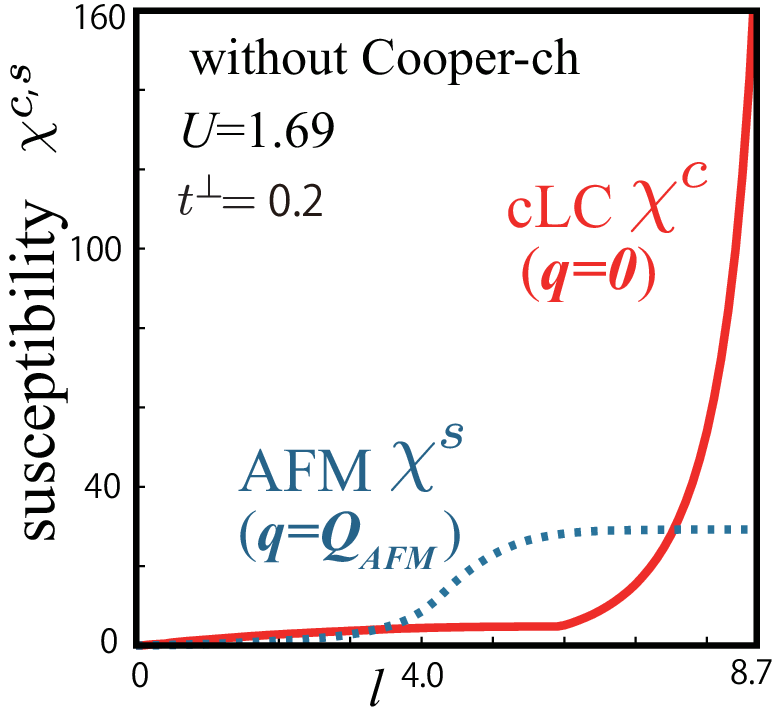}
\caption{Flows of the charge- and spin-ch susceptibilities without contribution
from the Cooper-ch scattering. here, $U=1.69,t^{\perp}$ and $T=0.05$.}
\label{fig:ppnashi}
\end{figure}

In the main text, we explain that the 
Peierls-ch scatterings give the dominant contribution to the cLC,
while both Peierls- and Cooper-ch scatterings are taken into account on an
equal footing in the fRG method.
To verify the important of the Peierls-ch for the cLC, we solve the RG equation without Cooper-ch, which is given by
\begin{eqnarray}\hspace{-5pt}
\frac{d g_{p_1p_2p_3p_4}}{d\Lambda_l}\hspace{-3pt}
=\sum_{pp'}
 \frac{d W_{\bm{p,p'}}^{+}}{d\Lambda_l}
\Bigl( g_{p_1p_3pp'} g_{pp'p_2p_4} \hspace{-5pt}
-g_{p_1p_2pp'} g_{pp'p_3p_4} \Bigr). \nonumber \label{RGeq_noC}
\end{eqnarray} 
In Fig. \ref{fig:ppnashi}, we plot the maximum value of
the charge (spin)-ch susceptibilities $\chi^{c(s)}(\equiv \max_{\q} \chi^{c(s)}(\q))$
without Cooper-ch contribution.
The charge-ch fluctuations develop at $\q=\bm{0}$ even if we drop the Cooper-ch.
The obtained result is quite similar as the one in the main text with Cooper-ch
as shown in Fig.\ref{fig:phase} (b).
Therefore, we conclude that particle-hole (AFM) fluctuations
play a significant role on the microscopic mechanism of the cLC,
while the $d$SC fluctuations are irrelevant.
\begin{figure}[htb]
 \begin{center}
\includegraphics[width=.99\linewidth]{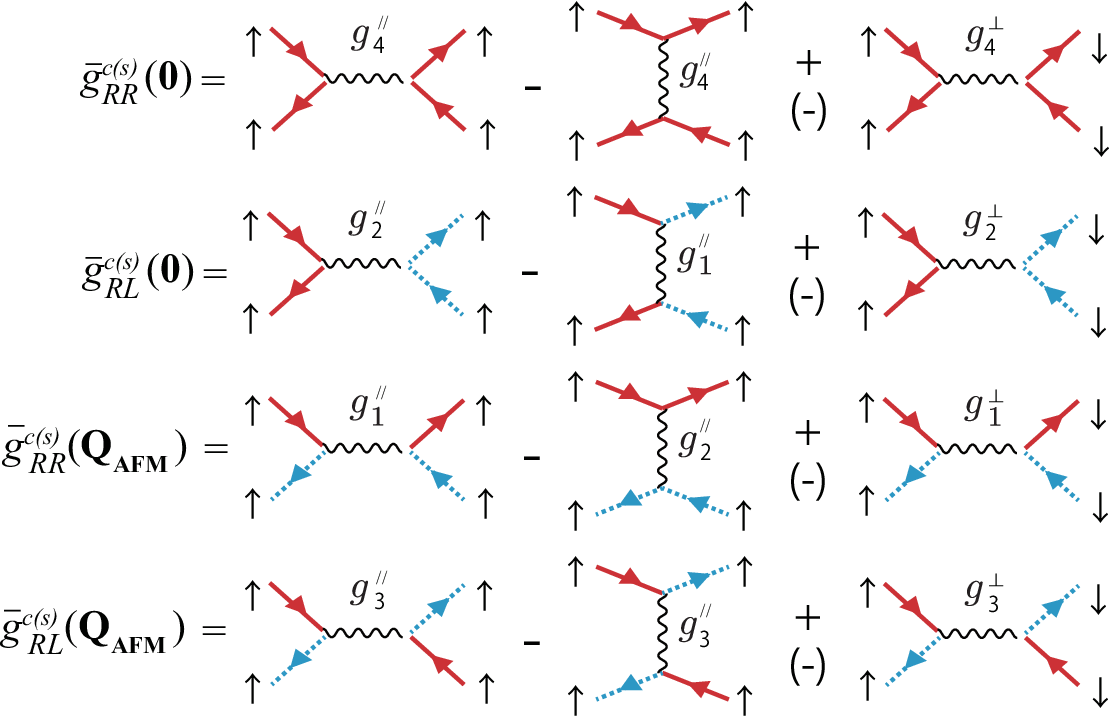}
\caption{One-to-one correspondence
between  $\bar{g}_{p p'}^{c(s)}(\q)$ and $g_{i}$.}
\label{fig:gology_def}
 \end{center}
\end{figure}
\subsection{D. $g$-ology theory at finite temperatures}
Here, we summarize the $g$-ology analysis for the cLC 
on the present q1D Hubbard model 
with linear energy dispersion.
\begin{figure*}[htb]
 \begin{center}
\includegraphics[width=.95\linewidth]{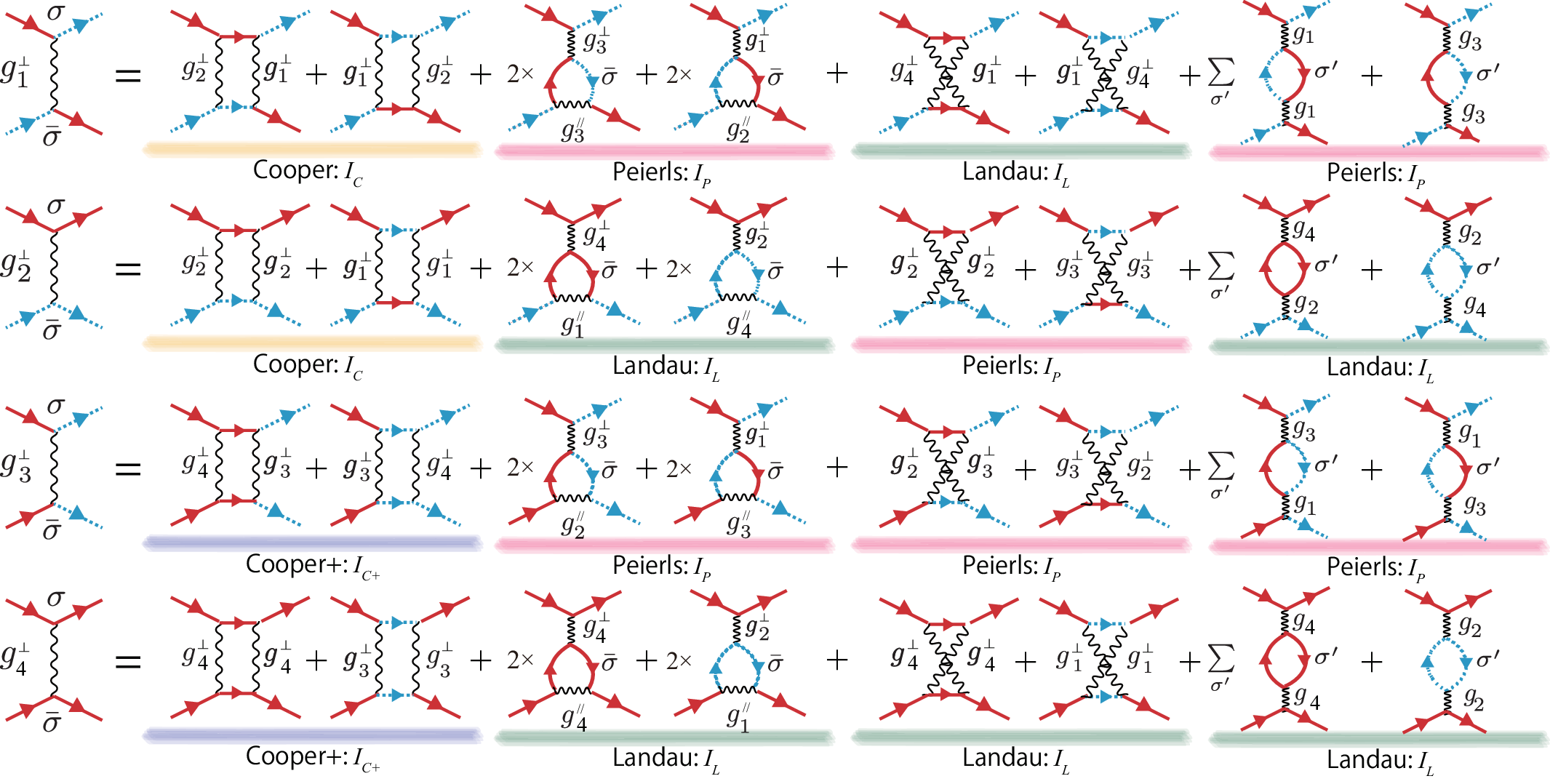}
\caption{The RG equation for the 4-point vertex $g_{i}$ in the 1-loop approximation.}
\label{fig:long}
 \end{center}
\end{figure*}
In the $g$-ology theory
\cite{SEmery,SBourbonnais,SKishine1,SKishine4,
SSuzumura,SSuzumura2,SSolyom}, the 4-point vertex function is
classified into 4-types; backward ($g_1$), forward ($g_2,g_4$) and umklapp ($g_3$) 
scatterings, which are defined in Fig. \ref{fig:golog}(c) in the main text.
By using the $g_{i} (i=1-4)$, the charge- (spin-) ch 4-point vertex 
$\bar{g}^{c}_{p,p'}(\bm{0})$ in the q1D model is
given by 
\begin{eqnarray}
&\!\!\!\!\!\! \bar{g}^{c}_{RR}(\bm{0})\approx 2\pi v_F g_4, \hspace{3pt}
&\!\!\! \bar{g}^{c}_{LR}(\bm{0})\approx 2\pi v_F(2g_2 -g_1) \nonumber \\
&\!\!\!\!\!\! \bar{g}^{s}_{RR}(\bm{Q}_{\rm{AFM}})\approx -2\pi v_F g_2, \hspace{3pt}
&\!\!\! \bar{g}^{s}_{LR}(\bm{Q}_{\rm{AFM}})\approx  -2\pi v_F g_3 
\label{eq:golo1_sap},
\end{eqnarray}
where its derivation is given in Fig. \ref{fig:gology_def}.
Based on these relations, we discuss the origin of
the cLC in terms of the $g$-ology theory in the main text.
The results of the present study are given in the TABLE I.
\begin{center}
\begin{table}
\begin{tabular}{|c|c|c|c|} \hline
$\hat{g}$ in q1D & $g$-ology & initial & present study \\ \hline \hline
$\bar{g}^{c}_{{RR}}(\bm{0})$ & $2\pi v_F\cdot g_4$ & $U$& large negative \\ \hline
$\bar{g}^{c}_{{LR}}(\bm{0})$ & $2\pi v_F(2g_2-g_1)$ & $U$& large positive\\  \hline
$-\bar{g}^{s}_{{RR}}(\bm{Q}_{\rm AFM})$ & $2\pi v_F\cdot g_2$ & $U$  & large positive \\  \hline
$-\bar{g}^{s}_{{LR}}(\bm{Q}_{\rm AFM})$ & $2\pi v_F\cdot g_3$  & $U$ & positive \\  \hline
\end{tabular}
\caption{(1st column) The charge- and spin-ch 4-point vertexes,
(2nd column) $g_{i}$ in the $g$-ology theory, (3rd column) initial value,
(4th column) the present results.} \label{tab:tab1}
\end{table}
\end{center}

The most important fact is that $g_{2}$ ($g_{4}$) is renormalized to large positive (negative)
value in the present q1D model as given in Fig. \ref{fig:golog}(d) in the main text.
To understand the large positive (negative) value of $g_2 \ (g_4)$, we consider the
1-loop RG equation for $g_{i}$. We note that $g_{i}$ has spin dependence
and is classified into 2-ch in the spin-conserving system: $g_{i}^{\perp}$ and $g_{i}^{\parallel}$.
$g_{i}^{\perp}$ is the interaction between fermions 
with the anti-parallel spin,
while $g_{i}^{\parallel}$ is that with the same spin.
By using the  $g_{i}^{\perp}$ and $g_{i}^{\parallel}$, the RG equation is given by
\begin{eqnarray}
\frac{dg_{1}^{\perp}}{dl}&=&2g_1^{\perp} g_2^{\perp} I_{C}+2g_1^{\perp}( g_2^{\parallel}-g_1^{\parallel}) I_{P}+2g_1^{\perp} g_4^{\perp} I_{L}, \nonumber \\ 
\frac{dg_2^{\perp}}{dl}&=&(g_2^{\perp}g_2^{\perp} +g_1^{\perp}g_1^{\perp}) I_{C} +2 g_4^{\perp}(g_1^{\parallel}-g_2^{\parallel}) I_{L}\nonumber \\ &+&
(g_2^{\perp} g_2^{\perp}  +g_3^{\perp} g_3^{\perp} ) I_{P}, \nonumber  \\
\frac{dg_3^{\perp}}{dl}&=&2g_3^{\perp} g_4^{\perp} I_{C+}+2g_3^{\perp}(g_2^{\parallel}-
g_1^{\parallel})I_{P}+2g_3^{\perp}g_2^{\perp}I_{P},\nonumber 
\\
\frac{dg_4^{\perp}}{dl}&=&(g_4^{\perp} g_4^{\perp}+g_3^{\perp}g_3^{\perp}) I_{C+}\nonumber \\ &+&
2g_2^{\perp}(g_1^{\parallel}-g_2^{\parallel})I_{L}+(g_4^{\perp} g_4^{\perp}+g_1^{\perp}g_1^{\perp}) I_{L}. \label{eq:goloperp}
\end{eqnarray}
Their diagrammatic expressions are given in Fig. \ref{fig:long}.
\begin{figure}[htb]
\includegraphics[width=.55\linewidth]{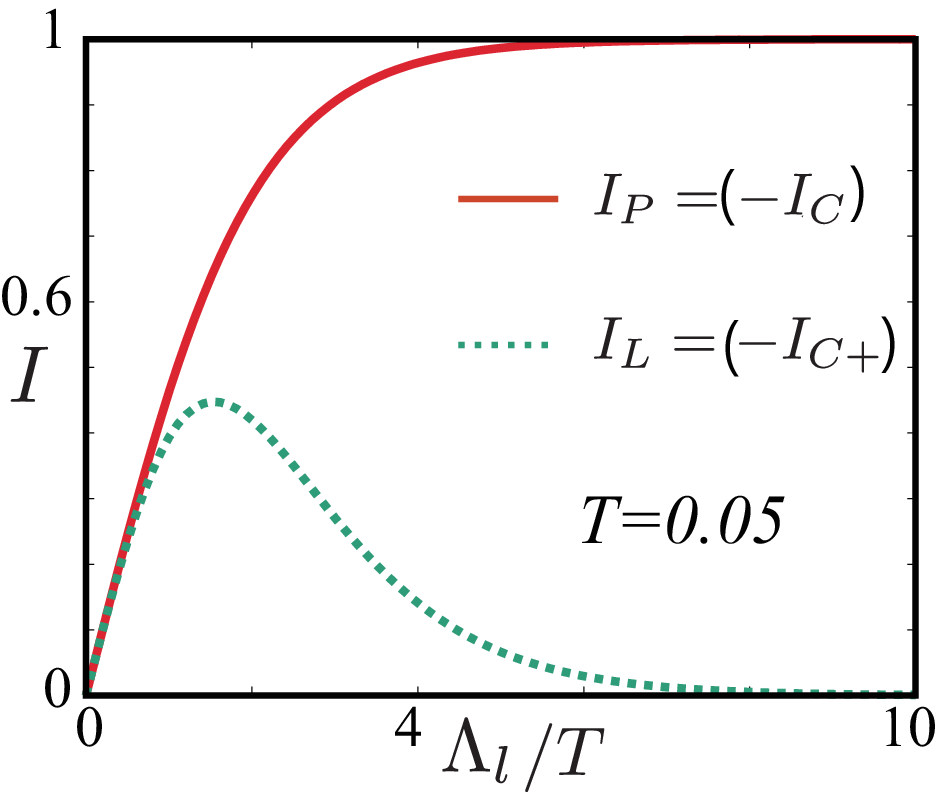}
\caption{$\Lambda_{l}$-dependence of the 
$I_{P}=-I_{C}= \tanh(\Lambda_l/2T)$ (red solid line) 
and $I_{L}=-I_{C+}=(\Lambda_l/2T)\cosh^{-2}(\Lambda_l/2T)$ (green dotted line).}
\label{fig:ppnashi2}
\end{figure}
Here, $I_P \ (I_L)$ corresponds to the Peierls (Landau) ch scatterings
defined by $I_P \ (I_L) \equiv 2\pi v_F \cdot dW_{\bm{R},\bm{L} (\bm{R},\bm{R})}^{+}/dl$.
On the other hand, $I_{C(C+)}$ is the Cooper (Cooper+) ch ones where
$I_C \ (I_{C+}) \equiv 2\pi v_F \cdot dW_{\bm{R},\bm{L} (\bm{R},\bm{R})}^{-}/dl$.

At finite temperatures,
The $\Lambda_{l}/T$-dependence of $I_{P}=-I_{C}$ and 
$I_{L}=-I_{C+}$ in the linear dispersion model
are shown in Fig. \ref{fig:ppnashi2}.
(At $T=0$, $I_P=-I_{C}=1$ and $I_L=I_{C+}=0$.)
The Landau-ch scattering 
plays important roles in the lower energy region comparable to
Cooper-ch (Peierls-ch) scattering \cite{Sfuseya_g4}.
Note that $I_{L} = 2\Lambda_l [L(\k,\q=\bm{0})]_{\e_\k=\Lambda_l}$ and 
$I_{C} = - 2\Lambda_l[C(\k)]_{\e_\k=\Lambda_l}$
in Fig. \ref{fig:sc}(c) in the main text.
We comment that the Landau and Cooper+ channels are 
completely dropped if we set $T=0$ before solving the RG equation. 
In this case, $g_4$ is marginal since $dg_4/dl=0$.
Therefore, we have to solve the RG equation at finite $T$,
and then we can take the $T\rightarrow0$ limit at the final stage.

By using the SU(2)-symmetry in the spin space, the relation $g_{1}^{\perp}-g_{2}^{\perp}=g_1^{\parallel}-g_2^{\parallel}$ holds. Thus, all of the $g^{\parallel}$ terms in Eq.(\ref{eq:goloperp}) are
replaced by $g^{\perp}$. 
Note that the $g^{\parallel}_{3}$ and $g^{\parallel}_{4}$ does not affect the 
physical quantity due to the anti-commutation relation of the fermion.
Thus, the expression $g_{i}$ in the main text stands for $g_{i}^{\perp}$. 
In addition, the Cooper-ch scattering is negligible as discussed in SM. C.
Then, the RG equations for $g_2$ and $g_4$ are simply rewritten as
\begin{eqnarray}
\frac{dg_2}{dl}&=&(g^2_2 +g^2_3) I_{P} +(-2g_2 g_4+2g_1g_4) I_{L},
\nonumber \\
\frac{dg_4}{dl}&=& (g^2_4 +g^2_1-2g^2_2 +2g_1g_2) I_{L},
\label{eq:goloperp2}
\end{eqnarray}
where we put $g_{i} \equiv g_{i}^{\perp}$. 
Therefore, $g_2$ is enhanced by the Peierls-ch scattering by
$(g^2_2 +g^2_3) I_{P}$, while it is suppressed by $-2g_2 g_4 I_{L}$.
(Note that $g_1$ is marginally irrelevant.)
On the other hand, $g_4$ is renormalized by the Landau-ch scattering.
In particular, $g_4$ becomes negative value due to the factor $-2g^2_2 I_{L}$.
Therefore, we conclude that the cLC is understood by the
important roles of $g_2$ and $g_4$, which are renormalized by Peierls- and Landau-ch
scatterings, respectively.

\subsection{E. Analysis of the fRG without frustration}
\begin{figure}[htb]
\includegraphics[width=.99\linewidth]{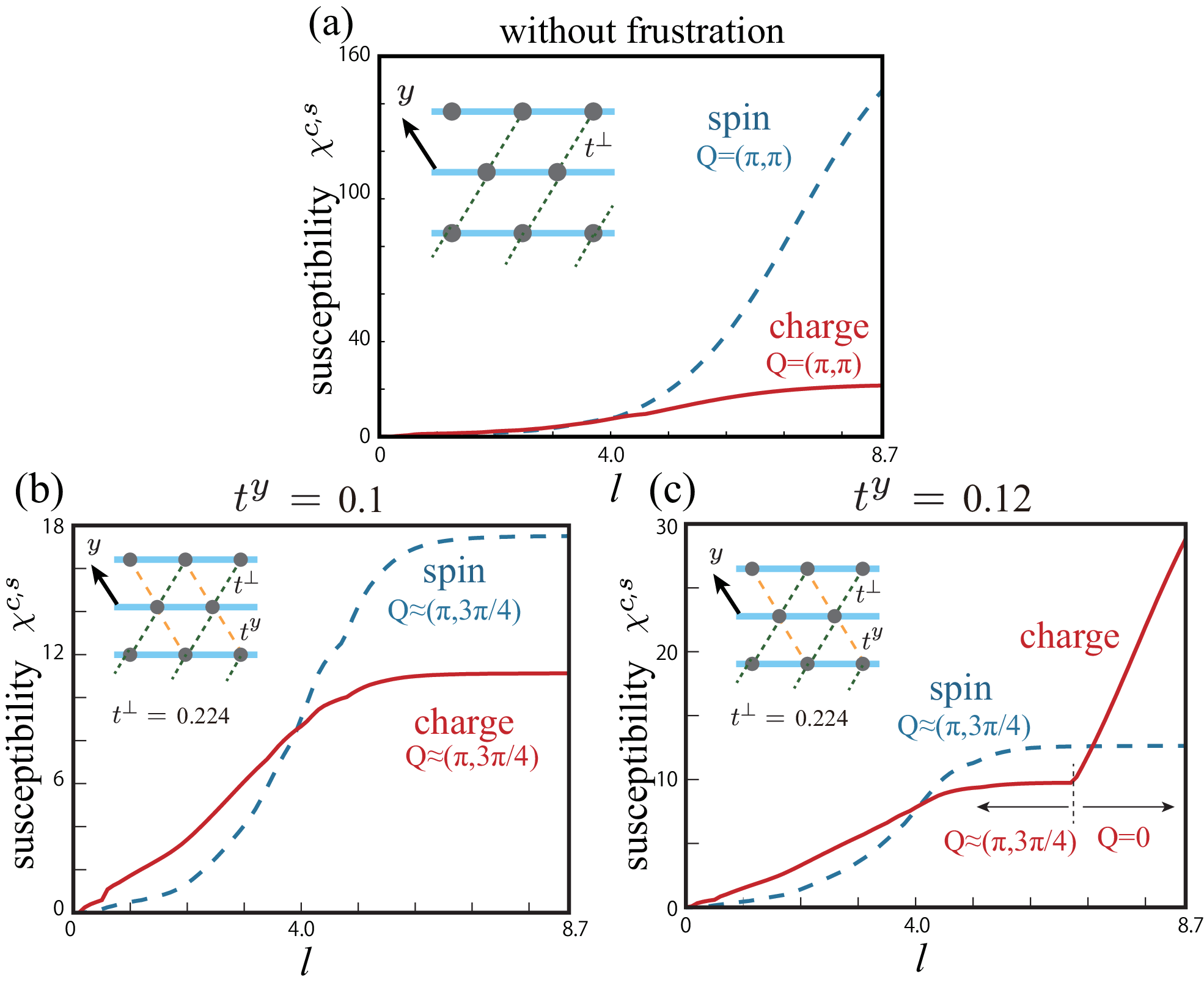}
\caption{(a) Flows of the maximum values for the charge- and spin-ch susceptibilities without the geometrical frustration at $t^{y}=0$, $U=1.56$ and  $\bm{Q}=(\pi,\pi)$. We also show the flows with the weak geometrical frustration at (b) $t^{y}=0.1$ and (c) $t^{y}=0.12$.} 
\label{fig:nofrust}
\end{figure}
Here, we discuss the importance of the magnetic frustration for the
emergence of the cLC.
For this purpose, we perform the fRG on a
coupled Hubbard chain model without frustration,
which is realized by cutting the hopping along the $y$-axis; $t^{y}$.
Figure \ref{fig:nofrust} (a) shows the obtained charge- and spin-ch susceptibilities
at $U=1.56, t^{y}=0, t^{\perp}=0.224$. The value of $U$ is slightly smaller than that in the main text by reflecting the absence of the magnetic frustration.
Only spin-ch fluctuations develop at $\bm{Q}=(\pi,\pi)$ while the charge-ch fluctuations
remain small. Here, the charge-ch fluctuations
correspond to the bond-order at $\bm{Q}=(\pi,\pi)$.
The present AFM phase is located in the FL region,
since the magnetic transition temperature is $T=0.05 \ll t^{\perp}$.

Therefore, the cLC does not appear without the magnetic frustration
even in the FL region. We verified that the cLC order can overcome the AFM order for $t^y/t^\perp \gtrsim0.5$ for $t^\perp=0.224$ at $T=0.05$ as shown in Figs. \ref{fig:nofrust} (b) and (c).

In summary, the geometry frustration strongly suppresses the AFM susceptibility.
On the other hand, the AFM fluctuation is important in the present cLC mechanism.
Thus,  it is useful to verify the correlation between the cLC order and spin fluctuation strength experimentally in order to discriminate the present cLC mechanism from the other cLC ones. Thus, it is possible to discriminate the true cLC mechanism by performing careful multiple experiments.
This fact brings a change to develop the Landau-ch scattering that is the origin of the cLC.
Therefore, we conclude that the magnetic frustrations are essential ingredient
to obtain the cLC in the main text. 


\end{document}